\documentclass[final]{aipproc}

\def\lesssim{\mathrel{\hbox{\rlap{\hbox{\lower4pt\hbox{$\sim$}}}\hbox{$<$}}}}

\layoutstyle{6x9}

\begin{document}

\title{(Sort of) Testing relativity with extreme mass ratio inspirals}

\classification{04.25.Nx, 04.30.Db, 04.70.Bw}
\keywords{Black holes, gravitational waves}

\author{Scott A.\ Hughes}{
  address={Department of Physics and MIT Kavli Institute\\
  Massachusetts Institute of Technology, Cambridge, MA 02139} }

\begin{abstract}
The inspirals of ``small'' ($1 - 100\,M_\odot$) compact bodies through
highly relativistic orbits of massive (several $\times 10^5\,M_\odot
-$ several $\times 10^6\,M_\odot$) black holes are among the most
anticipated sources for the LISA gravitational-wave antenna.  The
measurement of these waves is expected to map the spacetime of the
larger body with high precision, allowing us to test in detail the
hypothesis that black hole candidates are described by the Kerr metric
of general relativity.  In this article, we will briefly describe how
these sources can be used to perform such a test.  These proposed
measurements are often described as ``testing relativity''.  This
description is at best somewhat glib: Because --- at least to date ---
all work related to these measurements assumes general relativity as
the theoretical framework in which these tests are performed, the
measurements cannot be said to ``test relativity'' in a fundamental
way.  More accurately, they test the {\it nature of massive compact
bodies within general relativity}.  A surprising result for such a
test could point to deviations from general relativity, and would
provide an experimentally motivated direction in which to pursue tests
of gravity theories beyond GR.
\end{abstract}

\maketitle

\section{Extreme mass ratio inspirals and their waves}

Compact binary sources in which one member is far more massive than
another produce {\it extreme mass ratio inspirals}, or EMRIs.
Binaries of this sort are expected to form by scattering processes in
the nuclei of galaxies: multibody interactions in the dense cluster of
stars surrounding a galaxy's central black hole will sometimes place a
compact object on an orbit that is so strongly bound to the black hole
that it is more likely to inspiral due to gravitational-wave (GW)
backreaction than it is to be scattered back out by further
interactions.  See Clovis Hopman's article in these proceedings
{\cite{clovis}} for a detailed description.

There has been quite an evolution in our thinking about these sources
over the past decade.  At the time of the 1st LISA Symposium, it was
thought that the inspiralling compact body was most likely to be a
white dwarf {\cite{sigrees,sig97}}; the inferred LISA event rate for
white-dwarf-dominated EMRIs was expected to be about an event per
year.  It is now believed that measured EMRIs will be dominated by the
capture of stellar mass ($\sim 10\,M_\odot$) black holes.  Though the
black hole birth rate is much lower than that of white dwarfs, black
holes are captured more efficiently due to {\it mass segregation} (the
tendency of heavier stellar objects to sink to the bottom of their
gravitational potential well) {\cite{spitzer}}.  In addition, black
hole EMRIs are ``audible'' to much greater distances since the EMRI
wavestrain scales with the mass of the smaller body.  These factors,
plus physics such as ``resonant relaxation'' (see Hopman's article
{\cite{clovis}} for discussion), have combined to boost the estimated
event rate to several hundred events per year!  Capture might in fact
be so efficient that galaxies ``eat'' all their EMRI sources early in
their lives, leaving few low-redshift EMRIs for LISA unless they are
replenished through later stellar evolution (a possibility discussed
by Sigurdsson and Rees {\cite{sigrees}}).  Conversely, the event rate
might be so large that {\it too many} EMRIs will be present in LISA
data, and the sources may be confused {\cite{bc04_conf}}.  GW data
will have a lot to say about the astrophysics of these sources.

The relativity community has been attracted to this source in large
part because it is an amazingly pristine astrophysical system.  Since
the mass ratio of EMRIs is, by definition, very small ($10^{-3}
\lesssim m_1/m_2 \lesssim 10^{-7}$), it can be treated as a
perturbative parameter.  These binaries can thus be analyzed using a
variant of perturbation theory, expanding the spacetime (or a suitable
surrogate for the spacetime) about the Kerr background of the binary's
larger member.  Schematically, the binary's evolution can be regarded
as a ``forced geodesic'': Letting $x^\alpha$ stand for the coordinates
of the smaller body, and writing the connection for the background
spacetime as $^{\rm B}{\Gamma^\alpha}_{\beta\gamma}$, the smaller body
follows a trajectory given by
\begin{equation}
\frac{d^2x^\alpha}{d\tau^2} + ^{\rm B}{\Gamma^\alpha}_{\beta\gamma}
\frac{dx^\beta}{d\tau}\frac{dx^\gamma}{d\tau} = f^\alpha_{\rm SF}\;.
\label{eq:schematic}
\end{equation}
If the right-hand side were zero, this would simply be the geodesic
equation describing motion in the background spacetime.  The {\it self
force} $f^\alpha_{\rm SF}$ that appears instead describes how the
motion is changed due to the small body's interaction with its own
distortion to the background spacetime.  The challenge becomes
computing this self force and its impact upon the small body's motion
{\cite{mst97,qw97}}, and then computing the GWs that this motion
generates.  See Poisson {\cite{ericLR}} for a particularly readable
review.  For our purposes, it suffices to note that computing the self
force without any additional simplifying assumptions (other than
extreme mass ratio) has proven to be quite a challenge.

Yasushi Mino {\cite{mino03}} demonstrated that {\it if} the evolution
can be regarded as ``adiabatic'' (in a sense to be quantified
momentarily), then the self force can be simplified to something that
is relatively tractable:
\begin{equation}
f^\alpha_{\rm SF} = \frac{1}{2}\left(f^\alpha_{\rm ret} -
f^\alpha_{\rm adv}\right)\;.
\end{equation}
(This result is identical in form to the flat space electromagnetic
self force originally derived by Dirac {\cite{dirac}}.)  In contrast
to the exact expression, which requires integrating a Green's function
over the entire past world line of the orbiting body, the ``retarded''
and ``advanced'' contributions are relatively simple to evaluate.  The
assumption of adiabaticity reduces that integration to an averaging
procedure; what remains can be used to describe the binary's evolution
using tools that are not much more complicated than those which have
been developed for simpler cases.  See {\cite{hdff,dfh,sthn}} for
further discussion.

The adiabatic assumption that makes it possible to use Mino's ``half
advanced minus half retarded'' prescription amounts to a separation of
time scales.  We require that the change in any quantity that
characterizes the orbit (such as energy or angular momentum) be
``small'' over a single ``orbit''.  Thus, over short timescales, we
can treat the system as following an orbit that neglects radiative
backreaction; over long timescales, we treat the inspiral as a
``flow'' of the system through a sequence of these orbits.

\begin{figure}
  \includegraphics[height=.4\textheight]{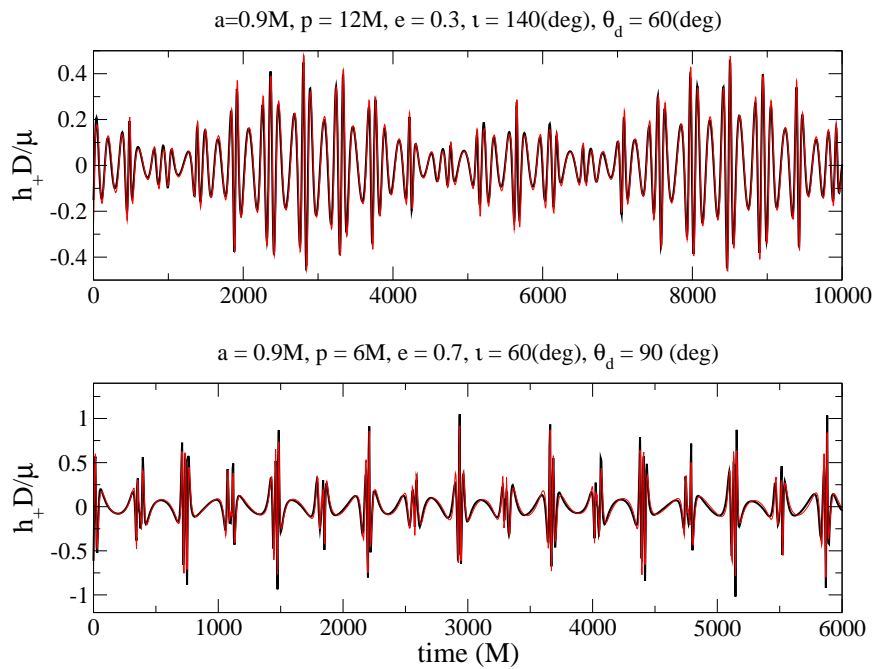}
  \caption{Generic EMRI waveforms.  Top panel shows waveform generated
  by a small body in a moderately strong field, moderately eccentric,
  highly inclined orbit about a rapidly spinning black hole; bottom
  panel shows waveform for a significantly eccentric and strong field
  orbit at shallower inclination angle about the same hole.  See Ref.\
  {\cite{dh06}} for detailed discussion.  Superplotted on these
  waveforms are the ``kludge'' waves described in Ref.\
  {\cite{kludge2}} (from which this plot is taken).  The kludge
  matches the strong-field calculation so well that the differences
  can barely be noted in this plot.}
  \label{fig1}
\end{figure}

\begin{figure}
  \includegraphics[height=.4\textheight]{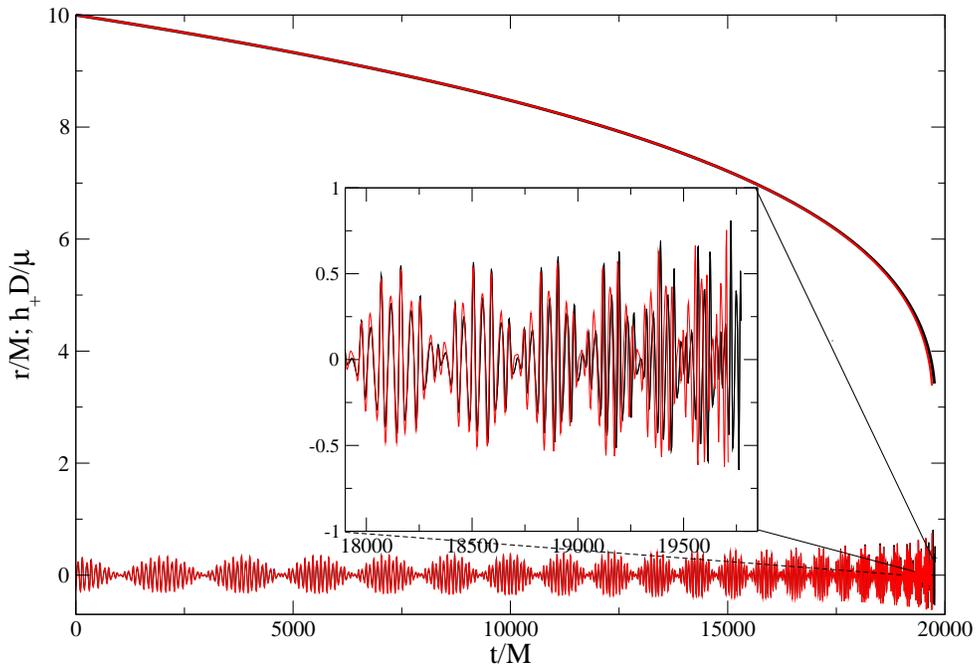}
  \caption{Circular EMRI inspiral waveform.  This plot shows waves in
  the (unrealistic) limit of zero eccentricity, but includes the
  effect of radiative backreaction; the upper curve shows the orbital
  radius as a function of time, $r(t)$.  (The waves shown in Fig.\
  {\ref{fig1}} are of sufficiently short duration that the effect of
  backreaction is neglected.)  This example is for inspiral into a
  rapidly rotating hole with spin parameter $a = 0.9$.  Superplotted
  on this waveform is a ``kludge'' inspiral as described in Ref.\
  {\cite{kludge2}} (from which this plot is taken).  As in Fig.\
  {\ref{fig1}}, the kludge matches the strong-field embarrassingly
  well.}
  \label{fig2}
\end{figure}

In practice, we currently treat the short timescale orbits as
geodesics of the background spacetime.  This construction throws away
``conservative'' components of the self force --- a time symmetric,
non-dissipative piece of $f^\alpha_{\rm SF}$ that pushes the orbit
away from a geodesic of the background spacetime even in the absence
of radiative backreaction.  Neglect of this piece will lead to a
systematic phase error in models of the waveforms {\cite{ppn05}}; this
will have to be fixed to maximize the quality of LISA measurements
(though it may be adequate for initial EMRI searches
{\cite{hf_inprep,fav_inprep}}).  Waveforms produced by this
approximation are being developed now {\cite{dh06}}, and appear to be
sufficiently accurate for developing and testing LISA EMRI data
analysis algorithms.  Some examples are shown in Figs.\ {\ref{fig1}}
and {\ref{fig2}}.

It's worth noting that recent work {\cite{kludge1,kludge2}} has shown
that almost all of the features of the waveforms developed in the
strong field analysis can be reproduced using a very simple ``kludge''
approximation in which a {\it flat spacetime} wave formula is coupled
to the geodesic motion of a small body orbiting a black hole.  Indeed,
the agreement between the two approaches is somewhat embarrassing ---
we have superplotted waveforms produced by the ``kludge'' on the
strong field results in Figs.\ {\ref{fig1}} and {\ref{fig2}}.  The
differences are very small; in particular, the phase coherency of the
two approaches is remarkable.  Since the ``kludge'' waves can be
computed {\it many} times quicker than the strong-field waveforms (cpu
minutes rather than cpu hours), we expect that the ``kludge'' approach
will play a key role in scoping out and developing LISA EMRI data
analysis.

\section{Mapping spacetimes and testing ``black-holey-ness''}

Comparing the EMRI waveforms that have been developed to the LISA
sensitivity shows that they can be tools for high precision studies of
black holes.  Finn and Thorne {\cite{ft2000}} showed that, as the
small body spirals through the strong field region of the larger black
hole, it executes tens to hundreds of thousands of orbits, taking on
the order of a year to do so.  Because so many orbits accumulate over
the measurement, one expects that a fit to a theoretical model will be
highly sensitive to small changes to the model's parameters, leading
to very accurate parameter determination.  Indeed, Barack and Cutler
{\cite{bc04_param}} demonstrated using simplified waveform models that
LISA should be able to determine many EMRI parameters with exquisite
accuracy.  In particular, the mass and spin of the larger black hole
should be determined with fractional errors of about $10^{-3} -
10^{-5}$ (depending upon the detailed nature of the particular EMRI).

Barack and Cutler's analysis assumes that the background spacetime is
a Kerr black hole.  That such amazing precision is possible under this
assumption begs the question of how well we could do if we assume more
general spacetimes.  In other words, rather than just measuring the
properties of the assumed black hole, can we test whether the large
central object is in fact a black hole, or is something even more
exotic?

Fintan Ryan {\cite{ryan1}} first formulated this question in a way
suitable for testing with GW observations.  The basic idea is to note
that the exterior spacetime of {\it any} stationary, axisymmetric body
is fully specified by a pair of multipole moment families
{\cite{geroch,hansen}}: a set of mass moments $M_l$, and a set of
current moments $S_l$.  These moments label the rate at which certain
fields constructed from the metric fall off with distance, as well as
their angular dependence.  (For a material body, these moments are
closely related to the ``usual'' multipolar distribution of mass and
mass motions in the body's interior {\cite{thorne80}}.)  As such, the
orbits and the GWs generated from the orbits are set by the multipole
moments that determine a spacetime's structure.  Turning this around,
it should be possible to infer the multipole moments which
characterize a spacetime by measuring the GWs generated by an object
orbiting in that spacetime.

Such a construction is closely analogous to the science of {\it
geodesy} --- inferring the distribution of the earth's mass by
studying the orbits of satellites.  It is a particularly powerful test
for black holes because they have a {\it very} special multipole
moment structure.  Given a Kerr black hole with mass $M$ and spin
parameter $a$ (using units in which $G = 1 = c$), the moments $M_l$
and $S_l$ are given by {\cite{hansen}}
\begin{equation}
M_l + i S_l = M(ia)^l\;.
\label{eq:kerr_mult}
\end{equation}
In other words, $M_0 = M$, $S_1 = a M$.  {\it All other multipoles are
completely determined by these two values, or else are zero.}

We thus have a very simple and natural consistency test that the
central object must satisfy if it is a Kerr black hole: Measure at
least three multipoles.  Numbers 0 and 1 fully determine numbers 2 and
higher.  {\it If any inconsistency with the Kerr relationship
(\ref{eq:kerr_mult}) is found, then that object cannot be a black
hole.}  This proposed measurement has a similar ``moral'' foundation
to the multimode spectroscopy presented at this meeting by Berti, and
discussed by Dreyer et al {\cite{dkkfgla}} and by Berti, Cardoso, and
Will {\cite{bcw}}: the goal is to {\it overdetermine} the parameter
space to check consistency with Kerr solution.

An initial exploration by Ryan {\cite{ryan2}} indicates that ``black
hole geodesy'' (which has been named ``bothrodesy'', based on the
Greek root $\beta o\theta\!\rho o\varsigma$, referring to a
sacrificial pit; and ``holiodesy'', based on the word ``hole'' and
making a pun on ``heliodesy'') should easily work well enough to test
the black hole hypothesis --- at least three moments should typically
be measured by LISA with better than a few percent accuracy (depending
on the large object's mass).  As the multipole order is increased,
measurement accuracy degrades.  Ryan's analysis indicates that as many
as 6 or 7 moments may be determined.  (Indeed, Ryan's analysis is
probably somewhat pessimistic, since for simplicity he only considers
circular, equatorial orbits.  The additional information encoded by
the multiply periodic structure of generic orbits is sure to sharpen
our ability to determine a spacetime's geometry.)

Despite this good news, Ryan's analysis can only be considered a proof
of principle --- as implemented, this spacetime mapping via multipoles
is unlikely to work for astrophysical massive objects.  In order to
make this ``assume nothing, measure all multipoles {\it ab initio}''
scheme work, Ryan must build a spacetime with {\it arbitrary}
multipole moments.  To make a tractable calculation, Ryan essentially
abused a weak-field approximation: He built a spacetime that was
strictly accurate only for $r \gg M$ (where $M$ is the mass of the
large central object), but used it for all $r$ (including $r \sim
\mbox{a few}\times M$).  In principle one could iterate the algorithm
Ryan used to build a spacetime with arbitrary multipoles good for all
$r$.  In practice, this is unlikely to work well due to the many
poorly convergent sums that arise from such an iteration.  Effective
bothrodesy is likely to require us to set our sights a little lower
than an assumption-free measurement of all of a massive body's
gravitational multipole moments.

It is often the case that the most effective measurements are of
quantities whose values, in some particular framework, should be zero.
Motivated by this, we have suggested reformulating bothrodesy as a
{\it null experiment}.  Rather than trying to measure some arbitrary
set of multipoles $\{M_l,S_l\}$, measure their deviation from the
values we would expect {\it if} the spacetime were a black hole:
$\{\delta M_l, \delta S_l\} \equiv \{M_l - M_l^{\rm BH}, S_l -
S_l^{\rm BH}\}$.  We have given the name {\it bumpy black hole} to
spacetimes in which the multipole deviations $\{\delta M_l,\delta
S_l\}$ are non zero, since such objects typically have an event
horizon, but one that is distorted from that of the Kerr metric
{\cite{ch04}}.

The bumpy black hole is a nice construction since it works very well
into the strong field, and also includes the black hole limit in a
natural way --- simply set the ``bumpiness parameters'' which control
$\{\delta M_l,\delta S_l\}$ to zero and we recover the ``normal''
black hole solutions.  Unfortunately, they have a somewhat
pathological feature: Bumpy black hole spacetimes contain naked
singularities.  In retrospect, this is not too surprising: Price's
theorem tells us that, in any ``normal'' situation, a small deviation
from a black hole solution should quickly radiate away.  Circumventing
this mechanism requires some mechanism to prevent normal physics from
acting; we are almost guaranteed that these solutions will not be ones
that we would encounter in nature.  We should bear in mind, though,
that the goal of this construction is not to build a spacetime that
might exist in nature.  Instead, we only want to provide a
well-behaved {\it falsifiable} straw man with which to test the black
hole hypothesis.  In this context, it is irrelevant whether deviations
from ``black holey-ness'' come from reasonable physics.  (Indeed, we
were recently reminded {\cite{st_private}} that the idea that a
``pure'' multipole is not necessarily physical is not unique to
gravity.  A similar issue applies to electromagnetic multipoles ---
when we model the electromagnetic field of an object by a sum over
multipoles, we don't actually believe that the object contains
infinitesimal loops of infinite current, but we accept this fiction
for the purpose of computing the object's external fields.  The
problem is magnified for gravity, however, due to the nonlinearity of
the field equations.)

The original presentation of this variant of bothrodesy focused, for
simplicity on angular-momentum-free spacetimes (i.e., bumpy
Schwarzschild holes) {\cite{ch04}}.  Since then, Glampedakis and Babak
have extended this idea to encorporate spinning black holes
{\cite{gb06}}.  Their ``quasi-Kerr'' spacetimes amount to Kerr black
holes with the ``wrong'' quadrupole moments.  (Though they do not
interpret the horizon structure, if any, of these objects, we will for
brevity's sake refer to them as ``bumpy black holes'' since their
motivations and structure are sufficiently similar to the original
proposal {\cite{ch04}}.)  Though much work remains to compute the GWs
that would arise from inspiral in these spacetimes, and also to
ascertain whether this construction is ready for mapping the
spacetimes of massive compact objects, it is encouraging that the
theoretical foundations for this measurement have largely been set.

\section{Is this testing relativity?}

Bumpy black holes assume general relativity in their construction.
Thus, strictly speaking, using this framework to map the spacetimes of
massive compact objects {\it is not a test of general relativity}.  It
is instead a test of the structure of massive bodies {\it within}
general relativity.

It is, however, an extraordinarily precise test of that structure.
Although work remains to quantify exactly how well these measurements
can do, it seems very likely that the null hypothesis --- and thus the
multipole deviations will be zero, that these bodies {\it are}
described by the Kerr solution --- can be tested quite rigorously.
General relativity provides such a rigid framework that any measured
deviation will require explanation.  The explanation may turn out to
be prosaic --- e.g., systematic observation of the ``wrong''
quadrupole moment may be due to tidal interactions with the black
hole's local environment.  If not, it may point the way to a new
understanding of strong field gravity.

Tather than worrying about whether these measurements are truly
testing general relativity or not, we advocate remaining agnostic.
General relativity has, to date, been such a formidibly complete and
successful theory that it is hard to guess how it might require
modification.  Rather than looking for wholesale massive deviations in
the blueprint, mapping the spacetimes of massive bodies allows us to
precisely check the lay of our theory's foundations.  In doing so, we
also gain a wealth of precise astrophysical information on black hole
masses and spins, and on dynamical processes in the cores of galaxies.

\begin{theacknowledgments}
Our work on EMRI science and measurement has benefitted enormously
from discussions and collaborations with Stas Babak, Nathan Collins,
Steve Drasco, Hua Fang, \'Eanna Flanagan, Sam Finn, Joel Franklin,
Jonathan Gair, Kostas Glampedakis, Daniel Kennefick, Eric Poisson,
Saul Teukolsky, and Kip Thorne.  Support for this work specifically is
provided by NASA Grants NAG5-12906 and NNG05G1056; work on
gravitational-wave science generally is supported by NSF Grants
PHY-0244424 and PHY-0449884.  We also gratefully acknowledge support
from MIT's Class of 1956 Career Development Fund.
\end{theacknowledgments}


\bibliographystyle{aipprocl}

\end{document}